# Connect the dots… finding all possible orbits between two points

**Philip R Blanco**

Department of Astronomy, San Diego State University, San Diego, CA, United States of America

E-mail: pblanco@sdsu.edu

## Abstract

You have a satellite (spacecraft or asteroid) that moves under the gravitational influence of a massive central body and follows a Keplerian orbit around it (ellipse, parabola, or hyperbola). Given measurements of two positions in its orbit, what is the family of possible orbital paths that connects them? I use the conic section orbit's semi-latus rectum, directly related to orbital angular momentum, to parameterise these orbits. The solutions have applications to orbit determination, ballistic missiles, interplanetary interception, and targeted re-entry. I also show how they can be applied to solve the *Lambert problem* of finding the unique transfer orbit that connects two points in a specified time interval. These results are accessible to advanced undergraduate students in physics or aerospace engineering. Supplementary materials are provided online.

Supplementary material for this article is available online

Keywords: orbital motion, ballistics, energy conservation, angular momentum, spaceflight, Kepler's laws, Lambert problem

## 1. Introduction

Newton's universal law of gravitation predicts that the orbital path of a satellite around a massive spherical central body is a conic section (ellipse, parabola, or hyperbola) with the centre of attraction located at the focus closest to *periapsis*, the point of closest approach. In planar polar coordinates, the distance $r$ of the satellite from the centre of attraction is,

$$r = \frac{p}{1 + e \cos(\theta - \theta_0)}, \tag{1}$$

where $\theta_0$ is the angular coordinate of periapsis, $p$ is the semi-latus rectum, and $e$ the eccentricity [1].

A triumph of classical physics was to show that orbits of isolated planets and comets around the Sun follow equation (1) as a consequence of inverse-square law gravitation. This still leaves the problem of fitting orbits to observed positions in space, which was tackled after Newton's death by Gauss, Laplace, and Gibbs [2]. While the general problem of orbit determination using only Earth-based observations of measured positions on the sky is complicated, the simpler problem of finding all orbits that connect two points—a 2-point boundary value problem—is still instructive and has applications to orbital transfers, interception, and inter-continental ballistic missiles (ICBMs).

Here I present a solution to this problem based on geometrical parameters that are straightforward for students to visualise, and can be derived from an orbit's conserved angular momentum and mechanical energy. In section 2, I present standard astrodynamics results that relate these quantities to an orbit's geometry and the orbiter's instantaneous velocity components. . In section 3, I provide the equations for the family of orbits passing through two points in space, using the semi-latus rectum $p$ (which is directly related to the orbital angular momentum) as a variable parameter. In section 4, I apply these results to select orbits of interest. In section 5 I provide a method to calculate the transfer time between the two points, which I use to solve the *Lambert problem*, for which I recommend a numerical integration technique over traditional analytical schemes. Section 6 provides suggestions for further investigations.

## 2. Orbit properties from conserved quantities

The system consists of an orbiter of mass $m$ and an attracting body of mass $M$. Henceforth I assume that $m \ll M$ such that the attracting central body remains effectively stationary, so we ascribe the energy and angular momentum of the system to the orbiter. Newton showed that the shape of the orbit depends on the conserved physical quantities: the specific mechanical energy $\varepsilon$ and specific angular momentum $h$ of the system. Here, 'specific' means that we have divided out the mass $m$ of the orbiter so that

$$\varepsilon = \frac{1}{2}(v_r^2 + v_\theta^2) - \frac{GM}{r} \text{ and } h = rv_\theta = r^2 \frac{d\theta}{dt}, \tag{2}$$

where $r$ is the orbiter's distance from the centre of the attracting body of mass $M$, and $v_r$ and $v_\theta$ are the respective radial and azimuthal components of its velocity.

### 2.1. Relating orbital geometry to conserved quantities

Equation (1) provides insight into the form of the orbit: $p$ is the 'half-width', $e$ is the 'squish', while $\theta_0$ is the 'twist'. See figure 1. Introductory texts and papers on orbital mechanics [1–3] derive the relationship between these geometrical properties and conserved physical quantities. The semi-latus rectum of the conic section is related to the specific orbital angular momentum by

$$p = a(1-e^2) = \frac{h^2}{GM}, \quad (3)$$

where $a$ is the semi-major axis of the orbit ('half-length' in the elliptical case, see figure 1). Since from equation (1) the orbital distance $r = p$ at true anomalies $\theta - \theta_0 = \pm 90°$, it is easy to compare the specific angular momenta of two orbital paths by comparing their latus rectum lengths, drawing a line passing through the nearside focus at right angles to the major axis, as shown in figure 1.

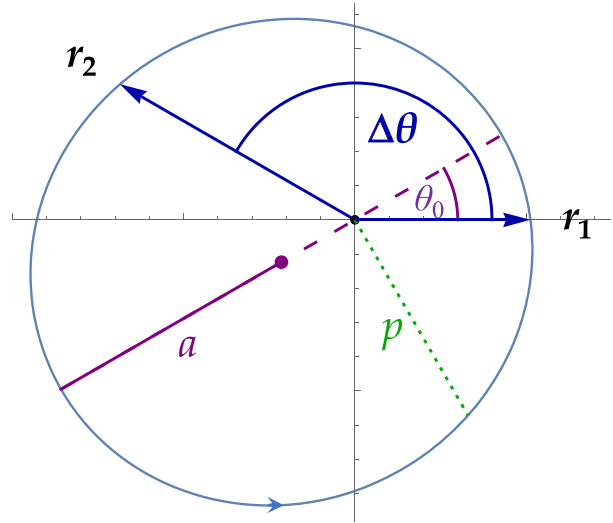

**Figure 1.** Geometry of an orbit connecting two points at respective polar coordinates $(r_1, 0)$ and $(r_2, \Delta\theta)$, with the centre of attraction at the origin. The angular coordinate of periapsis is $\theta_0$. The geometrical centre of the ellipse is marked with a solid circle; the bold line between it and apoapsis is the semi-major axis of the ellipse, length $a$. The dotted line is the semi-latus rectum, length $p$.

Another standard result [1–3] relates the semi-major axis of an orbit to the specific mechanical energy,

$$a = \frac{-GM}{2\varepsilon} \quad (4)$$

for all nonparabolic orbits with $\varepsilon \neq 0$. For bound elliptical orbits, this allows us to compare the specific energy of two orbits by measuring their major axes—the larger ellipse has a higher (but still negative) specific energy.

The 'squish' or orbital eccentricity $e \geqslant 0$ of the orbit can then be expressed in terms of these conserved quantities as,

$$e = \sqrt{1 - \frac{p}{a}} = \sqrt{1 + \frac{2\varepsilon h^2}{(GM)^2}}. \quad (5)$$

For non-zero angular momentum $h$, $e = 0$ describes a circular orbit, $0 < e < 1$ a bound elliptical orbit, $e = 1$ a parabola (corresponding to a minimally unbound $\varepsilon = 0$) and $e > 1$ a hyperbolic escape (on the branch for which $|\theta - \theta_0| < \arccos(-1/e)$, from equation (1)).

## 2.2. Velocity components in orbit

Once a desired orbital path is chosen, it is useful to determine the azimuthal and radial velocity components $v_\theta$ and $v_r$ required to achieve that orbit. We shall assume all orbits are prograde with angle $\theta$ increasing with time, such that $v_\theta > 0$. The azimuthal velocity is obtained from the specific orbital angular momentum, or the semi-latus rectum, via equations (2) and (3),

$$v_\theta = \frac{h}{r} = \frac{\sqrt{GMp}}{r} = \sqrt{\frac{GM}{p}}(1 + e\cos(\theta - \theta_0)). \quad (6)$$

To obtain the radial velocity component $v_r$ at any position, differentiate equation (1) to find the rate of change of orbital distance with angle, then use conservation of specific angular momentum in equations (2) and (3) and apply the chain rule,

$$v_r = \frac{dr}{dt} = \frac{d\theta}{dt} \cdot \frac{dr}{d\theta} = \frac{h}{r^2} \cdot \frac{dr}{d\theta} = \frac{\sqrt{GMp}}{r^2} \frac{pe\sin(\theta - \theta_0)}{(1 + e\cos(\theta - \theta_0))^2}$$
$$= \sqrt{\frac{GM}{p}} e \sin(\theta - \theta_0). \quad (7)$$

The orbiter's radial velocity has the same sign as $\sin(\theta_0 - \theta_0)$, which makes sense when one inspects the example orbit of figure 1. Equation (7) also shows that the radial velocity has its maximum magnitude at $\theta - \theta_0 = \pm 90°$ where $r = p$, the semi-latus rectum distance (where the dotted line intersects the orbit in figure 1). Moreover, since $\sin(\theta - \theta_0) = -\sin(\theta - \theta_0 + 180°)$, opposite points in the orbit have equal and opposite radial velocities. At the apsides of the orbit, $\theta - \theta_0 = 0°$ (periapsis) or $180°$ (apoapsis) and $v_r = 0$.

One can invert these relationships to find the orbital elements in equation (1) in terms of the velocity components at a given starting position $(r_1, 0)$. It is convenient to normalise the velocity components in terms of the local circular speed $v_C = \sqrt{GM/r_1}$ such that $v_\theta = c_\theta v_C$ and $v_r = c_r v_C$. Then from equation (2), the conserved quantities can be rewritten as $\varepsilon = \frac{1}{2}v_C^2(c_r^2 + c_\theta^2 - 2)$ and $h = r\, v_\theta\, v_C$. From equations (1), (3), (5), and (7) with $\theta = 0$, the orbital elements are,

$$p = r_1 c_\theta^2,\ e = \sqrt{1 + c_\theta^2(c_\theta^2 + c_r^2 - 2)},\ \theta_0 = \arctan\left(\frac{-c_r c_\theta}{c_\theta^2 - 1}\right), \quad (8)$$

with due regard to the numerator and denominator to select the correct quadrant for $\theta_0$. (Some treatments combine $e$ and $\theta_0$ to form the eccentricity vector $\vec{e} = (c_\theta^2 - 1)\hat{r} - c_r c_\theta \hat{\theta}$ that points in the direction of periapsis and has magnitude $e$; this is the other conserved vector quantity in an orbit [4]).

Equation (8) enables the reader to plot the entire orbit based on the orbiter's instantaneous position and velocity. Set $\theta = 0$ at this position. Then in terms of the velocity components $c_\theta$ and $c_r$ at $(r_1, 0)$ the orbit equation (1) becomes,

$$r(\theta) = \frac{r_1 c_\theta^2}{1 + \sqrt{1 + c_\theta^2(c_\theta^2 + c_r^2 - 2)} \cos\left[\theta - \arctan\left(\frac{-c_r c_\theta}{c_\theta^2 - 1}\right)\right]}$$

$$= \frac{r_1 c_\theta^2}{1 + (c_\theta^2 - 1)\cos\theta - c_\theta c_r \sin\theta}, \tag{9}$$

which is a less intuitive form, but will be used in section 3.2.

## 3. Finding all orbits connecting two points

Some students may be familiar with one kind of orbital transfer—the Hohmann transfer that uses purely azimuthal velocity changes to move an orbiter from one circular orbit to another [1, 2, 5]. If one's goal is simply to reach a given disance $r_2$ from the centre of attraction, starting in a circular orbit at $r_1$, the Hohmann transfer is usually optimal, with the first azimuthal impulse placing $r_1$ and $r_2$ at the apsides of the transfer orbit ($\theta = \theta_0$ and $\Delta\theta = 180°$ in figure 1).

A more general problem in astrodynamics is to find a desired orbital path between two aribitrary positions relative to the central body, with an angular separation $\Delta\theta \neq 180°$. We place the first position at polar coordinates $(r_1, 0)$ and the second at $(r_2, \Delta\theta)$ with $\Delta\theta > 0$ in the plane whose normal is given by the cross product of the position vectors. More general orbits in three dimensions can be transformed into such an anticlockwise (prograde) orbit in this plane.

### 3.1. The parable of the parabola

Consider projectile motion in a uniform gravitational field of strength $g$, where the goal is to launch a projectile from the origin to a final impact point $(x_f, y_f)$. Students learn that these boundary conditions constrain the relation between launch angle and speed, but an equally valid way to describe these paths is to choose a horizontal launch velocity $v_{x0}$, then adjust the vertical component $v_{y0}$ to hit the target. (Imagine a cart that rolls on a horizontal track at fixed speed $v_{x0}$, with a vertical launcher mounted on it such that the projectile speed $v_{y0}$ can be adjusted.) Once $v_{x0}$ is chosen, $v_{y0}$ is uniquely defined by it and the boundary conditions, as is the time of flight $t_f$,

$$x_f = v_{x0} t_f, \quad y_f = v_{y0} t_f - \tfrac{1}{2} g t_f^2$$
$$\Rightarrow v_{y0} = \frac{y_f}{x_f} v_{x0} + \frac{g x_f}{2 v_{x0}} \text{ with } t_f = \frac{x_f}{v_{x0}}. \tag{10}$$

We can take a similar approach to finding all possible orbits that connect two positions in space, by creating a family of connecting orbits where the radial velocity $c_r$ is a function of azimuthal velocity $c_\theta$. Once $c_\theta$ is chosen, $c_r$ and the transfer time between the points take on unique values.

### 3.2. A family of orbits connecting two points

Equation (9) represents an orbital path passing through $(r_1, 0)$. If it subsequently passes through $(r_2, \Delta\theta)$ it must satisfy

$$r_2 = \frac{r_1 c_\theta^2}{1 + (c_\theta^2 - 1)\cos\Delta\theta - c_\theta c_r \sin\Delta\theta}$$

$$\Rightarrow c_r = \frac{(c_\theta^2 - 1)\cos\Delta\theta + 1 - \frac{r_1}{r_2}c_\theta^2}{c_\theta \sin\Delta\theta}. \tag{11}$$

Similar to the projectile motion case, the boundary conditions $r_1$, $r_2$, and $\Delta\theta$ constrain the normalised radial velocity $c_r$ to be a single-valued function of azimuthal $c_\theta$. In the limit of small $\Delta\theta \approx x_f/r$ and $y_f \ll r$ on a planet's surface with $v_C = \sqrt{gr}$, equation (11) agrees with equation (10); I provide a detailed proof in the supplementary materials.

To draw the orbit, use either equation (9) or equations (1) and (8) to define the geometrical parameters $p$, $\theta_0$, and $e$. Hereon I use the semi-latus rectum $p = r_1 c_\theta^2$ as the independent variable, since it is easier to visualise (as a 'half-width') than the azimuthal velocity component $c_\theta$.

The periapsis angle is found from equations (8) and (11),

$$\theta_0 = \arctan\left[\frac{r_1(p - r_2)\csc\Delta\theta - r_2(p - r_1)\cot\Delta\theta}{r_2(p - r_1)}\right], \tag{12}$$

with due regard to the signs of numerator and denominator to select the correct quadrant. $\theta_0$ becomes undefined when $p = r_1 = r_2$ for a circular orbit with no unique periapsis.

The eccentricity can be then be found either from the orbit equation (1) for the first point, or equations (8) and (11),

$$\begin{aligned} e &= (p/r_1 - 1)\sec\theta_0 \\ &= \sqrt{\frac{\left(\frac{p}{r_1} - \frac{p}{r_2}\right)^2}{2(1 - \cos\Delta\theta)} + \frac{\left(\frac{p}{r_1} + \frac{p}{r_2} - 2\right)^2}{2(1 + \cos\Delta\theta)}}, \end{aligned} \tag{13}$$

where the second expression shows that the solution for $e$ is symmetrical to the interchange of $r_1$ and $r_2$.

Equations (11)–(13) fail for $\Delta\theta = 180°$ (which includes the Hohmann transfer [2, 5]) where the two known points are diametrically opposite the centre of attraction. This is because $c_\theta$ and hence semi-latus rectum $p$ take on a single value at $\Delta\theta = 180°$, and $c_r$ becomes a free parameter. I cover that case in the online supplement. This configuration of points is also undesirable in 3-dimensional space because it does not define the orbital plane, and so should be avoided where another measurement of $(r_2, \Delta\theta \neq 180°)$ is possible.

In the following sections, I use equations (1), (12), and (13) in preference to equations (9) and (11), since they provide insights into the orbital geometry. Unlike the projectile motion example of section 3.1, the time of flight (transfer time) between positions in an orbit is more complicated to calculate, and is discussed in section 5.

### 3.3. An illustrative example

Consider two points with $r_2 = {}^5/_3 r_1$ separated by $\Delta\theta = 120°$, as shown in figure 2. A realistic example in Earth orbit could have $r_1 = 9000$ km and $r_2 = 15\,000$ km. Alternatively, $r_1 = 1$ AU and $r_2 = 1.67$ AU approximates an interplanetary transfer between the heliocentric orbits of Earth and Mars. Figure 2 shows some connecting orbits as solutions to equations (12) and (13) for different values of $p$.

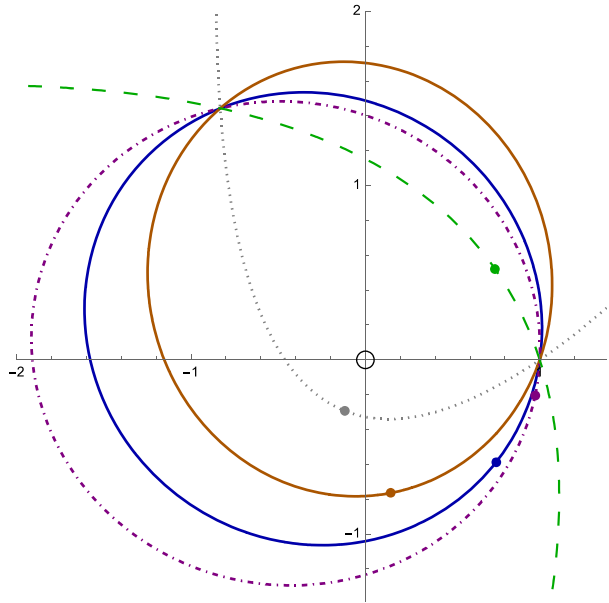

**Figure 2.** Some orbits connecting two points with $r_2 = {}^5/_3\, r_1$ and $\Delta\theta = 120°$, described by equations (12) and (13). The direction of motion is anticlockwise. The periapsis position is denoted by a dot in each case: Blue: Minimum eccentricity, $p = {}^{60}/_{49}\, r_1$, $e = {}^2/_7$, $\theta_0 = 321.8°$, section 4.1. Orange: Minimum orbital energy (and semi-major axis), $p = {}^{15}/_{14}\, r_1$, $e = 1/\sqrt{7}$, $\theta_0 = 280.9°$, section 4.2. Dot-dashed purple: Minimum-impulse from a circular orbit, $p = 1.3128\, r_1$, $e = 0.3194$, $\theta_0 = 348.3°$, section 4.3. Dashed green: Connecting parabolic trajectory, $p = p_{\mathrm{par},2} = 1.8173\, r_1$, $e \equiv 1$, $\theta_0 = +35.2°$, section 3.4. Dotted grey: Limiting parabolic trajectory (does not connect in the prograde direction), $p = p_{\mathrm{par},1} = 0.6317 r_1$, $e \equiv 1$, $\theta_0 = 248.4°$.

### 3.4. Constraints on p for a prograde connecting orbit

For the example in figure 2 with $\Delta\theta = 120°$, we see that for small values of $p$, the two points cannot be connected by a prograde unbound parabola or hyperbola, since $r \to \infty$ at some angle between 0 and $\Delta\theta$. The limiting cases of unbound parabolic orbits are found by setting $e = 1$ in equation (13). This gives the two semi-latus rectum values $p_{\mathrm{par},1}$ and $p_{\mathrm{par},2}$ corresponding to parabolic paths,

$$p_{\mathrm{par},1} = \frac{2r_1 r_2 \sin^2 \frac{\Delta\theta}{2}}{r_1 + r_2 + 2\sqrt{r_1 r_2} \cos \frac{\Delta\theta}{2}}, \text{ and}$$

$$p_{\mathrm{par},2} = \frac{2r_1 r_2 \sin^2 \frac{\Delta\theta}{2}}{r_1 + r_2 - 2\sqrt{r_1 r_2} \cos \frac{\Delta\theta}{2}} \quad (14)$$

with the sign of the cosine term determining which of these is the maximum and which is the minimum. It can be shown (most easily by plotting) that all valid prograde (anticlockwise) orbits connecting $r_1$ and $r_2$ have $p > p_{\mathrm{par},1}$ for $0 < \Delta\theta < 180°$, and $p < p_{\mathrm{par},1}$ for $180° < \Delta\theta < 360°$, with $p_{\text{par},2}$ providing the semi-latus rectum of the valid prograde connecting parabola. Figure 2 shows both parabolas for the example introduced in section 3.2.

These bounds on $p$ translate directly to bounds on the specific angular momentum $h$, from equation (3), a measure of 'motion around' the centre of attraction. Therefore, depending on the second point's position relative to the first, some orbits have either not enough (for $\Delta\theta < 180°$), or too much (for $\Delta\theta > 180°$) angular momentum to reach it.

## 4. Special cases and their applications

Equations (12) and (13) describe all orbits connecting two points as a function of the semi-latus rectum. The advantage of choosing the semi-latus rectum $p$ as a parameter is that every orbit with angular momentum, bound or unbound, has one; specifying $p$ uniquely defines the connecting prograde orbit, subject to the constraints discussed in section 3.4. In this section I present results for special cases that are commonly encountered in orbital mechanics.

### 4.1. Minimum eccentricity

The minimum eccentricity ('roundest') orbit connecting any two points is found from setting the derivative of equation (13) with respect to $p$ equal to zero, to give

$$p_{e,\,\min} = \frac{r_1 r_2 (r_1 + r_2)(1 - \cos\Delta\theta)}{r_1^2 + r_2^2 - 2r_1 r_2 \cos\Delta\theta}$$
$$= \frac{(r_1 + r_2)(d^2 - (r_1 - r_2)^2)}{2d^2} \text{ for } e_{\min} = \frac{|r_1 - r_2|}{d}, \quad (15)$$

where $d = \sqrt{r_1^2 + r_2^2 - 2r_1 r_2 \cos\Delta\theta}$ is the linear distance (or chord length) between the two positions in orbit (a line joining the tips of the position vectors in figure 1). The semi-major axis of this orbit is simply ½$(r_1 + r_2)$, from equation (3), which proves this orbit is always a bound ellipse. As expected, $e_{\min} = 0$ when $r_1 = r_2$, corresponding to a circular orbit with $p_{e,\min} = r_1 = r_2$. For the orbit described in section 3.2, equation (13) gives $p_{e,\min} = {}^{60}/{}_{49} r_1$ for $e_{\min} = {}^{2}/{}_{7}$, and $\theta_0 = 321.8°$ from equation (12). This orbit is shown in figure 2 as a solid blue ellipse.

The minimum-eccentricity orbit has some practical applications. Reducing variations in a satellite's distance and speed is advantageous for avoiding re-entry or impact, managing thermal effects, and reducing Doppler shift in communications. Non-Keplerian forces tend to 'circularise' the orbits of Earth-bound satellites. Therefore, equation (15) can serve as an initial estimate for $p$ when solving the Lambert problem described in section 5.

### 4.2. Minimum specific energy

Equation (4) relates orbital specific energy $\varepsilon$ to the orbit's semi-major axis $a$ via $\varepsilon = -GM/(2a)$, i.e. smaller orbits have lower energy. From equation (3) form $a = p/(1-e^2)$, then substitute from equations (12) and (13) and set the derivative of $a$ with respect to $p$ equal to zero. This gives a value for the smallest semi-major axis $a_{\min}$ that connects the two points, with

$$p_{a,\,\min} = \frac{d^2 - (r_1 - r_2)^2}{2d} \text{ for } a_{\min} = \frac{r_1 + r_2 + d}{4}$$

$$\Rightarrow e_{a,\,\min} = \sqrt{\frac{\frac{2}{d}(r_1 - r_2)^2 + r_1 + r_2 - d}{r_1 + r_2 + d}} \tag{16}$$

where the eccentricity is found from equation (13) or $e_{a,\min} = \sqrt{1 - p_{a,\min}/a_{\min}}$. The sum $r_1 + r_2 + d$ equals the perimeter of the triangle formed by the two orbital points and the centre of attraction, so $a_{\min} > 0$ always and the minimum-energy orbit is always a bound ellipse. From Kepler's third law, the orbital period is also minimised.

For the example orbit of section 3.2 with $r_2 = {}^5/_3 r_1$ and $\Delta\theta = 120°$, the minimum energy orbit has $p_{a,\min} = {}^{15}/_{14}\, r_1$, $a_{\min} = {}^5/_4\, r_1$, and $e = 1/\sqrt{7}$ from equation (16), with $\theta_0 = 280.9°$ from equation (12). This is the solid orange orbit in figure 2.

Since the minimum-energy orbit also has the smallest speed at $r_1$ from equation (2), it has an important application in ballistics. In a supplement I analyse cases with $r_2 = r_1$, which correspond to the exoatmospheric trajectories of ICBMs, where the launch speed to reach a target can be minimised by choosing the minimum-energy connecting orbit.

### 4.3. Impulsive transfer from a circular orbit

You have a satellite in a circular orbit at $(r_1, 0)$ and desire to execute an impulsive manoeuvre (a change in velocity with negligible change in position) to reach a position $(r_2, \Delta\theta)$ with $\Delta\theta > 0$. What is the speed change $\Delta v$ (also known as specific impulse) required for such a direct transfer from the circular orbit to the desired position? Applications include interplanetary targeting and directed re-entry.

From equation (6), since the satellite in a circular orbit is already moving with prograde velocity $v_C$, the change in azimuthal velocity for a given connecting orbit is given by,

$$\Delta v_\theta = v_\theta - v_C = v_C(c_\theta - 1) = v_C\left(\sqrt{\frac{p}{r_1}} - 1\right), \tag{17}$$

while equations (7) or (11) gives the radial velocity change at $\theta = 0$,

$$\Delta v_r = v_C(c_r - 0) = v_C \frac{r_2(p - r_1)\cos\Delta\theta - r_1(p - r_2)}{r_2\sqrt{pr_1}\sin\Delta\theta}, \tag{18}$$

after substituting $p = r_1 c_\theta^2$, and assuming $\Delta\theta \neq 180°$.

The total specific impulse is $\Delta v = \sqrt{\Delta v_r^2 + \Delta v_\theta^2}$. Figure 3 plots $\Delta v$ as a function of $p$ for the case $\Delta\theta = 120°$ and three values of the ratio $r_2/r_1$. If $r_2 = r_1$, the minimum $\Delta v = 0$ for $p = r_1$ and any $\Delta\theta$, since the orbiter can just continue its circular orbit. Otherwise, the value of $p$ at which $\Delta v$ is a minimum is the root of a quartic equation and best found by a numerical search. For the example of section 3.3, the minimum $\Delta v = 0.1563 v_C$ for $p = 1.3128 r_1$, shown as the dot-dashed purple orbit in figure 2. For the case $\Delta\theta = 180°$ the minimum $\Delta v$ is the first impulse of a Hohmann transfer [1, 5].

## 5. Transfer time and the Lambert problem

We saw in section 3.1 that for projectile motion in a uniform gravitational field, there is a family of trajectories that start and end at the same two points in space; the boundary conditions constrain the relationship between the components of launch velocity.

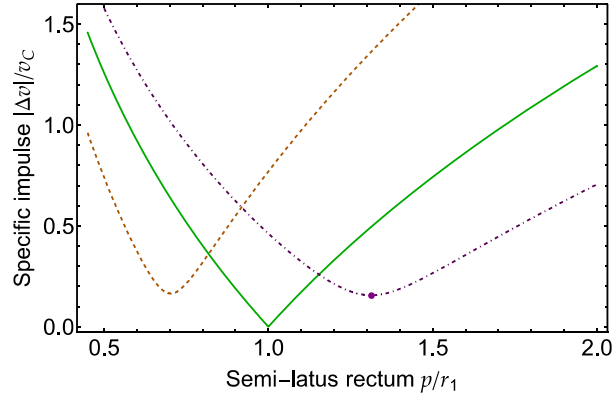

**Figure 3.** Specific impulse $\Delta v$ (relative to the circular orbital speed $v_C$ at $r_1$) to reach a given radius $r_2$ at $\Delta\theta = 120°$, from equations (17) and (18), as a function of the connecting orbit's semi-latus rectum $p$. Solid green curve: $r_2 = r_1$. Dashed orange curve: $r_2 = {}^3/_5\, r_1$. Dot-dashed purple curve: $r_2 = {}^5/_3\, r_1$, for which the minimum, marked with a dot, corresponds to the dot-dashed orbit in figure 2, with $p = 1.3128 r_1$ and $\Delta v = 0.1563\, v_C$.

Equation (10) shows that one can choose a unique trajectory from these paths by specifying either the horizontal launch speed or the time of flight.

The task of finding the unique orbit that connects two positions in a specified transfer time $\Delta t$ is known as the *Lambert problem* [2, 6]. Starting with conservation of angular momentum and the orbit equation (1), one can find the angular velocity $\omega$ at a point in the orbit,

$$h = r v_\theta = r^2 \frac{d\theta}{dt} = \sqrt{GMp}$$

$$\Rightarrow \omega = \frac{d\theta}{dt} = \frac{\sqrt{GMp}}{r^2} = \sqrt{\frac{GM}{p^3}} (1 + e \cos(\theta - \theta_0))^2. \quad (19)$$

The integral of this equation would give angular position versus time, locating the orbiter; however there is no closed-form analytic solution, and root-finding methods must be used to solve this *Kepler problem* [2, 3].

To find the transfer time between the two points needed for the Lambert problem, invert equation (19), then integrate over polar angle $\theta$ from 0 to $\Delta\theta$ to obtain,

$$\frac{dt}{d\theta} = \frac{r^2}{\sqrt{GMp}} = \sqrt{\frac{p^3}{GM}} \frac{1}{(1 + e \cos(\theta - \theta_0))^2}$$

$$\Rightarrow \Delta t = \frac{T_C}{2\pi} \left(\frac{p}{r_1}\right)^{\frac{3}{2}} \int_0^{\Delta\theta} (1 + e \cos(\theta - \theta_0))^{-2}\, d\theta, \quad (20)$$

where $T_C = 2\pi r_1/v_C$ is the circular orbital period at radius $r_1$, and equations (12) and (13) provide $\theta_0$ and $e$ as a function of $p$ for the connecting orbit.

While analytic solutions exist for this integral, the transcendental functions they contain depend on the values of orbital eccentricity: $0 < e < 1$ for elliptical, $e = 1$ for parabolic, and $e > 1$ for hyperbolic. Many textbooks derive the solutions using a geometrical method similar to that first described by Johannes Kepler himself [2, 3]. This requires the definition of new angular measures (mean and eccentric 'anomalies') to locate the object on its orbit, with extensions for the parabolic and hyperbolic cases. These methods were developed to allow for hand calculation of orbits and transfer times prior to digital computers.

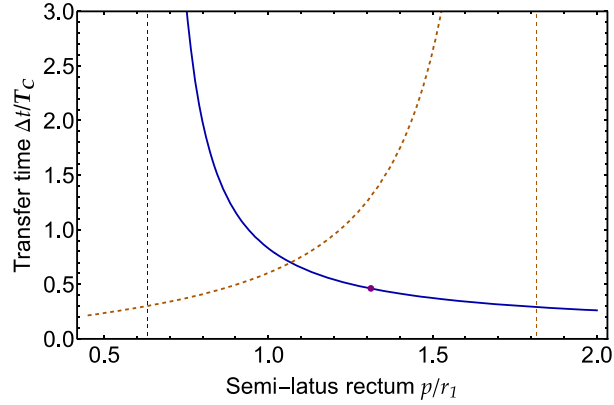

**Figure 4.** Transfer time $\Delta t$ (normalised by the circular orbital period $T_C$ at $r_1$), from $r_1$ to $r_2 = {}^5/_3 r_1$ at $\Delta\theta = 120°$ (blue curve) and $\Delta\theta = 240°$ (dashed orange curve), from equation (20). Vertical dashed lines show the values of $p/r_1$ corresponding to unbound parabolic orbits, from equation (11). A purple dot denotes the dot-dashed orbit in figure 2, $p = 1.3128 r_1$ and $\Delta t = 0.4607\, T_C$.

Instead, here I follow the recommendation of [6] to use numerical integration for equation (20), especially for student use. This works for all valid connecting orbits, regardless of eccentricity. Students can either use packaged numerical integrators or write their own—Simpson's rule integration is sufficient for this smooth integrand, and is used in the spreadsheet I provide in the supplementary materials.

One analytic solution is noteworthy: Setting $\Delta\theta = 360°$ for $e < 1$ gives the orbital period $T$, for which the integral of equation (20) evaluates to,

$$T = 2\pi \sqrt{\frac{p^3}{GM}} (1 - e^2)^{-3/2} = 2\pi \sqrt{\frac{a^3}{GM}}, \tag{21}$$

which is Newton's version of Kepler's 3rd law, where the period depends only on the semi-major axis.

As a practical example, use $GM = 3.986 \times 10^5 \text{ km}^3\text{ s}^{-2}$ for the Earth and take the values used in section 3.2, i.e. $r_1 = 9000$ km (for which $v_C = 6.655 \text{ km s}^{-1}$ and $T_C = 8497.2$ s) and $r_2 = {}^5/_3 r_1 = 15\,000$ km, separated by $\Delta\theta = 120°$. Connect these points with the minimum-$\Delta v$ orbit of section 4.3 that has $p = 1.3128 r_1 = 11816$ km, shown in figure 2. Equations (12) and (13) give $\theta_0 = 348.3°$ and $e = 0.3194$. Equation (20) gives a transfer time of $\Delta t = 0.4607 T_C = 3915$ s. The animation of figure 2 in the online supplementary materials shows this transfer.

Figure 4 shows $\Delta t$ values as a function of $p$ for $r_2 = {}^5/_3 r_1$ with $\Delta\theta = 120°$ as in the example above, and for $\Delta\theta = 240°$. For 'short way' transfers with $\Delta\theta < 180°$, $\Delta t$ decreases with $p > p_{\text{par},1}$ of equation (14), since the connecting orbit becomes faster at $r_1$ and 'straighter' (see figure 2). In the limit of a hyperbolic transfer, the connecting orbit becomes a straight line, and the transfer time is zero, but this requires an infinite orbital specific energy. Conversely, for 'long way' transfers with $\Delta\theta > 180°$, $\Delta t$ increases with $p < p_{\text{par},1}$ despite the increase in speed at $r_1$, since the connecting orbit gets larger with $p$, pushing apoapsis farther out, decreasing the angular velocity $\omega = h/r^2$ which increases the transfer time until it becomes infinite for an unbound parabola. Since $\Delta t$ is monotonic with $p$ in both cases, it is straightforward to find $p$ for the connecting orbit that has a desired transfer time, via a graphical (figure 4) or numerical search, solving the Lambert problem [2, 5, 6].

Time-specified conic trajectories of the Lambert problem are important for orbit determination from observations [7], interception and rendezvous such as targeting a spacecraft to reach a planet ($r_2 > r_1$) [8], or timed de-orbit ($r_2 < r_1$) [9]. A common application is to use the results of figures 3 and 4 together to find possible transfer times for a given available $\Delta v$ on board the spacecraft, or in reverse to find the necessary $\Delta v$ to achieve a desired transfer time.

## 6. Further investigations

### 6.1. Example problems

In the supplementary materials I provide example problems that can be solved with the equations and techniques presented here, most easily using the accompanying Excel spreadsheet. Similar problems can be found in advanced orbital mechanics texts [2], which use methods that are more difficult for students to follow, with additional steps and definitions of variables that have no obvious physical interpretation.

Here, one example will suffice to show the advanced level of analysis that students can achieve with the techniques described in this paper. This is a classic Lambert probem: A satellite was observed to make a coplanar transfer from a low Earth circular orbit ($r_1 = 7000$ km, $T_C = 97.14$ min)) to intercept the geostationary belt ($r_2 = 42\,000$ km), with $\Delta \theta = 165°$ in 300 min. What were the orbital elements for this transfer orbit, and the $\Delta v$ of the impulse? (Answer: $p = 11\,893$ km, $\theta_0 = 347.0°$, $e = 0.7173$, and $\Delta v = 2.472$ km s$^{-1}$).

### 6.2. Fitting an orbit to three points

A related problem is to fit an orbit through 3 points, which uniquely constrain $p$, $e$, and $\theta_0$ in equation (1) without needing time information. An analytic solution using vector algebra in 3 dimensions can be derived using a method pioneered by Gibbs [2], but the 2-point solution here can be applied as well, by finding all valid orbits (as a function of $p$) connecting any 2 points with $\Delta \theta \neq 180°$ from equations (12) and (13), then rearranging equation (1) to solve for the unique value of $p = r_3 (1 + e(p) \cos(\theta_3 - \theta_0(p)))$ to fit the third point located at polar coordinates $(r_3, \theta_3)$. Since I could not find the 3-point planar solution for semi-latus rectum $p$ in the literature, and it was excruciating to derive, I provide it here,

$$p = \frac{r_1 r_2 r_3 (\sin \theta_3 + \sin(\theta_2 - \theta_3) - \sin \theta_2)}{(r_1 r_3 \sin \theta_3 + r_2 r_3 \sin(\theta_2 - \theta_3) - r_1 r_2 \sin \theta_2)}, \tag{22}$$

with equations (12) and (13) yielding $\theta_0$ and $e$ as before, using any two points with $\Delta \theta \neq 180°$ and this value of $p$. (For clarity, here I use $\theta_2 \equiv \Delta \theta$ such that the three known points have polar coordinates $(r_1, 0)$, $(r_2, \theta_2)$, and $(r_3, \theta_3)$.) While the first position is anchored at angular coordinate $\theta = 0$, exchanging points 2 and 3 has no effect on $p$. Values for $p$ that are $\leqslant 0$, undefined $0/0$, or infinite will arise if the three points cannot be connected by a Keplerian orbit (e.g. if they are collinear), and plotting the points and orbit solution is useful to verify the sense of the orbital motion, especially for unbound cases with $e \geqslant 1$ (see section 3.4).

### 6.3. Propagation of uncertainties

If the points to be connected have measurement uncertainties, one can perform a numerical search on $p$ using least-squares methods, which then provides uncertainties for the orbital parameters. A related problem for students to explore is orbit prediction: Given two points in an orbit with uncertainties, how well can we predict the location of the orbiter at a later time? Students will realise that 'short arcs' (in time and space) do not constrain orbits well. This introduces them to a simplified form of the propagation of uncertainties in asteroid orbits [7]. Fitted orbits are needed to predict whether near-Earth objects pose a threat to Earth. Examples include Apophis [10] and recently 2024YR$_4$, which on discovery was deemed to have a 1% probability of an Earth collision, until subsequent measurements allowed the refinement of its orbit [11].

## 7. Conclusions

If two points on an orbit are known, equations (12) and (13) provide the orbital parameters of all possible connecting orbits, as a function of the conic section's semi-latus rectum, which is directly related to the orbital angular momentum. These allow the user to choose a connecting orbit that minimises either eccentricity, orbital energy, or specific impulse. One can then apply equation (20) to solve the time-constrained direct transfer Lambert problem via a straightforward numerical search on $p$, which works for valid bound and unbound orbits and is easier for students to visualise than more advanced methods [2].

I hope that the techniques described here will inspire teachers and students to explore the rich field of *astrodynamics* (defined as 'the study of natural and artificial objects' motions in space') while appreciating the underlying physical principles.

In the online supplementary materials, I provide an analysis of the cases $\Delta\theta = 180°$, $r_1 = r_2$ for ICBMs, and an animation of orbits connecting the two points shown in figure 2. I also provide an Excel spreadsheet for calculating transfer orbit parameters and times as a function of $p$, with plots of the resulting orbital paths. This can be used as a Lambert solver for many textbook problems, and for some example student exercises that I have also included.

### Acknowledgments

Alas, I received no help or encouragement with this work, but I am thankful to my employer Grossmont College for approving a sabbatical that permitted me to *focus* on these orbits. I thank an anonymous reviewer for carefully reading the original manuscript.

### Data availability statement

All data that support the findings of this study are included within the article (and any supplementary files).

### ORCID iDs

Philip R Blanco  https://orcid.org/0000-0002-8051-1673